\documentclass[epj]{webofc}
\usepackage[varg]{txfonts}   
\usepackage{caption}
\usepackage{subcaption}
\wocname{EPJ Web of Conferences}
\woctitle{CONF12}
\newcommand{\cD}{{\cal D}}

\newcommand{\be}{\begin{equation}}
\newcommand{\beq}{\begin{equation}}
\newcommand{\ee}{\end{equation}}
\newcommand{\eeq}{\end{equation}}
\newcommand{\bea}{\begin{eqnarray}}
\newcommand{\eea}{\end{eqnarray}}

\newcommand{\lk}{\left(}

\newcommand{\il}{\int\limits}

\def\R{{\mathbb{R}}}

\newcommand{\vB}{\vec{B}}

\newcommand{\vA}{\vec{A}}
\newcommand{\vx}{\vec{x}}
\newcommand{\vy}{\vec{y}}
\newcommand{\vz}{\vec{z}}
\newcommand{\vD}{\vec{D}}

\newcommand{\vp}{\vec{p}}
\newcommand{\vq}{\vec{q}}

\newcommand{\ii}{\mathrm{i}}
\newcommand{\dd}{\mathrm{d}}
\newcommand{\tr}{\mathrm{tr}}

\renewcommand{\vec}[1]{\mbox{\boldmath$#1$\unboldmath}}

\begin{document}
\selectlanguage{english}
\title{Hamiltonian approach to QCD in Coulomb gauge at zero and finite temperature\footnote{Talk given by H.~Reinhardt at "XIIth Quark Confinement and the Hadron Spectrum", 28 August-4 September 2016, Thessaloniki, Greece.}}
%
%

\author{H.~Reinhardt\inst{1}\fnsep\thanks{\email{hugo.reinhardt@uni-tuebingen.de}} \and
	G.~Burgio\inst{1} \and D.~Campagnari\inst{1} \and E.~Ebadati\inst{1} \and J.~Heffner\inst{1} \and M.~Quandt\inst{1} \and P.~Vastag\inst{1} \and
        H.~Vogt\inst{1}
}

\institute{Universit\"at T\"ubingen \\
Institut f\"ur Theoretische Physik \\
Auf der Morgenstelle 14 \\
D-72076 T\"ubingen \\
Germany}

\abstract{I report on recent results obtained within the Hamiltonian approach to QCD in Coulomb gauge. By relating the Gribov confinement scenario to the center vortex picture of confinement it is shown that the Coulomb string tension is tied to the spatial string tension. For the quark sector a vacuum wave functional is used which results in variational equations which are free of ultraviolet divergences. The variational approach is extended to finite temperatures by compactifying a spatial dimension. For the chiral and deconfinement phase transition pseudo-critical temperatures of $170 \, \mathrm{MeV}$ and $198 \, \mathrm{MeV}$, respectively, are obtained.
}
\maketitle
\section{Introduction}
\label{intro}

Motivated by the fact that lattice calculations are still unable to describe the QCD phase diagram at large baryon density, in recent years functional methods have been developed to 
study QCD non-perturbatively in the continuum. These methods are based either on Dyson--Schwinger equations \cite{Fischer2006, Alkofer2001, Binosi2009, Watson2006, Watson2007, Watson2008} or functional renormalization group flow equations \cite{Pawlowski2007, Gies2012}, or they exploit the variational principle in either the Hamiltonian \cite{Feuchter2004, Feuchter2005} or covariant \cite{Quandt2013, Quandt2015} formulation of gauge theory.

In this talk I will review some recent results obtained within the Hamiltonian approach to QCD in Coulomb gauge. After a short introduction to the basic features of this approach I will present lattice calculations which show that the so-called Coulomb string tension is linked not to the temporal but to the spatial string tension. Thereby I will show that the Gribov--Zwanziger confinement scenario is related to the center vortex picture of confinement. I will then report on new variational calculations carried out for the quark sector of QCD. After that I will extend the Hamiltonian approach to QCD in Coulomb gauge to finite temperatures by compactifying a spatial dimension. Numerical results will be given for the chiral and dual quark condensate. Finally, I will give some outlook on future research within the Hamiltonian approach. 

\section{Variational Hamiltonian approach to Yang--Mills theory}\label{sectII}

\begin{figure}
\centering
\begin{subfigure}{0.45\textwidth}
\includegraphics[width=\textwidth,clip]{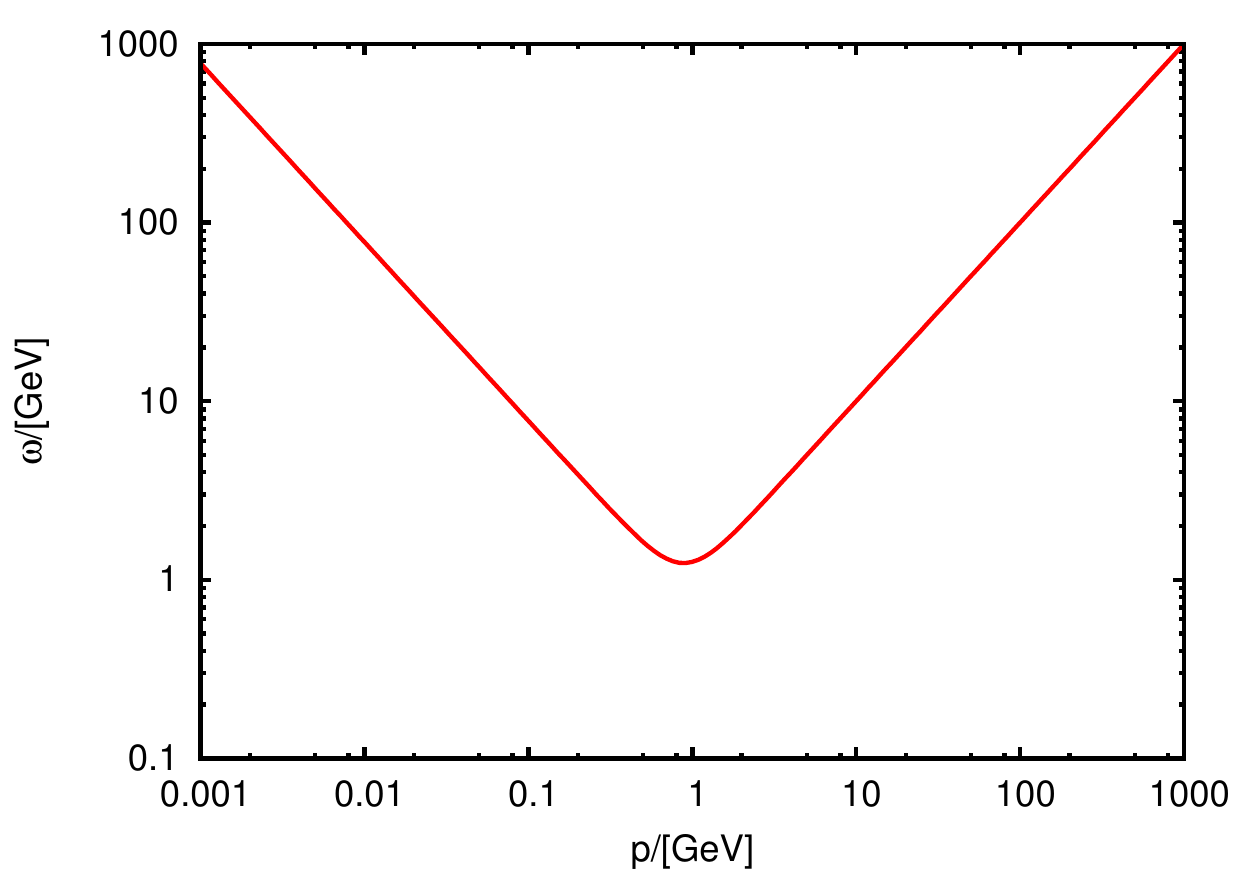}
\caption{}
\end{subfigure}
\quad
\begin{subfigure}{0.45\textwidth}
\includegraphics[width=\textwidth,clip]{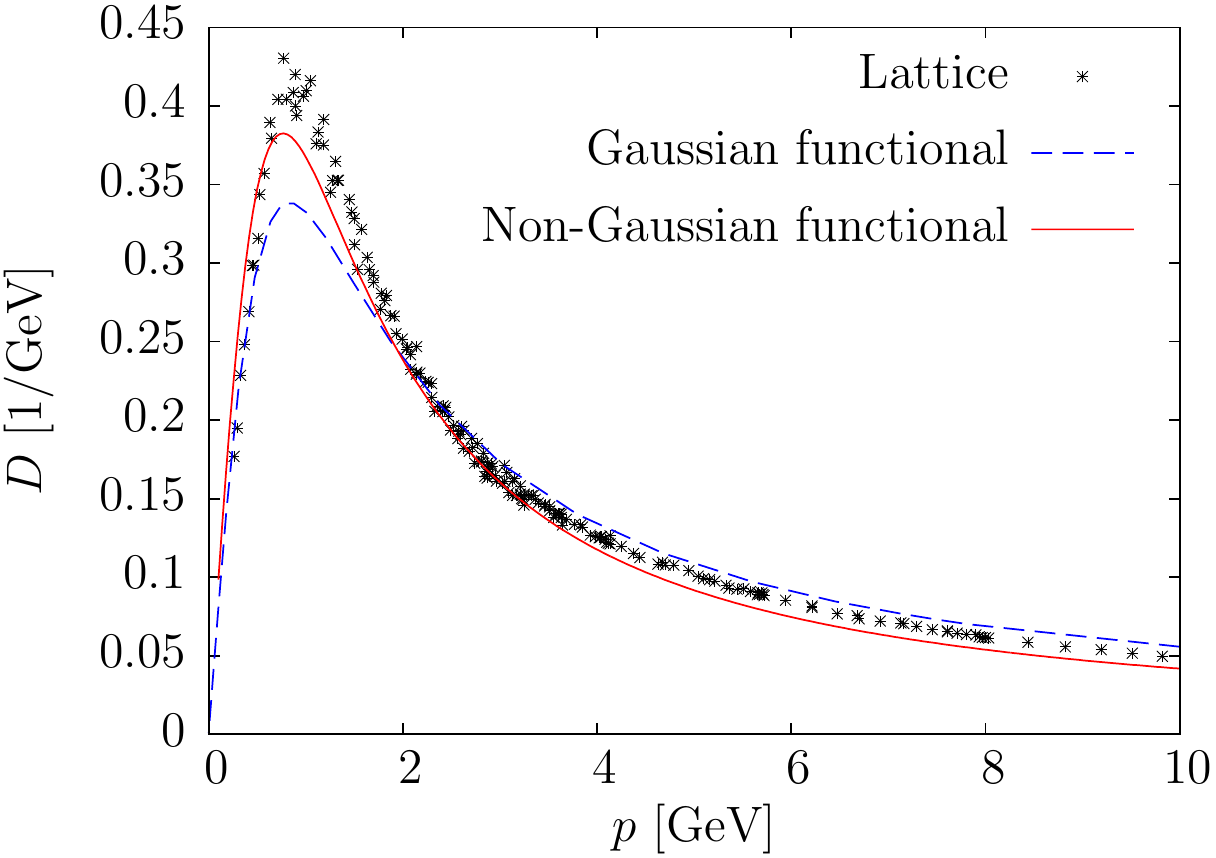}
\caption{}
\end{subfigure}
\caption{(a) Gluon energy $\omega$ obtained within the variational approach using the wave functional (\ref{419-G5}). (b) Gluon propagator $D = 1 / 2 \omega$ calculated on the lattice (crosses) compared to the results of variational calculations using a Gaussian ansatz, eq.~(\ref{419-G5}) (dashed curve), and a non-Gaussian ansatz, ref.~\cite{CR2010} (full curve).}
\label{fig1}%
\end{figure}

For pedagogical reason let me first summarize the basic features of the Hamiltonian approach in Coulomb gauge for pure Yang--Mills theory: After canonical quantization in Weyl gauge $A_0 = 0$ and resolution of Gau{\ss}'s law in Coulomb gauge $\vec{\partial} \cdot \vA = 0$ one finds the following gauge fixed Hamiltonian \cite{Christ1980}
\beq
H = H_{\mathrm{T}} + H_{\mathrm{C}}
\eeq
with
\beq
H_{\mathrm{T}} = \frac{1}{2} \int \dd^3 x \lk J^{-1}[A] \vec{\Pi}^a(\vx) \cdot J[A] \vec{\Pi}^a(\vx) + \vB^a(\vx) \cdot \vB^a(\vx) \right) \label{G1}
\eeq
where
\beq
B^a_k(\vx) = \varepsilon_{klm} \left(\partial_l A^a_m(\vx) - \frac{g}{2} f^{abc} A^b_l(\vx) A^c_m(\vx)\right)
\eeq
is the non-Abelian magnetic field with $g$ being the coupling constant, $\Pi^a_k (\vx) = \delta / (\ii \delta A^a_k (\vx))$ is the momentum operator (electric field) and $J[A] = \mathrm{Det}(-\hat{\vD} \cdot \vec{\partial})$ is the Fadeev--Popov determinant with $\hat{D}^{a b}_k (\vx) = \delta^{a b} \partial^x_k - g f^{a c b} A_k^c(\vx)$ being the covariant derivative in the adjoint representation of the gauge group. Furthermore,
\beq
H_{\mathrm{C}} = \frac{g^2}{2} \int \dd^3 x \int \dd^3 y \, J[A]^{-1} \rho^a(\vx) J[A] 
\left[(-\hat{\vD} \cdot \vec{\partial})^{-1} (-\vec{\partial}^2) (-\hat{\vD} 
\cdot \vec{\partial})^{-1}\right]^{a b}(\vx, \vy) \rho^b(\vy) \label{2}
\eeq
is the so-called Coulomb term with the color charge density
\beq
\rho^a(\vx) = f^{a b c} \vA^b(\vx) \cdot \vec{\Pi}^c(\vx) + \rho_m^a(\vx)\,. \label{3}
\eeq
This expression contains besides the charge density of the matter fields $\rho^a_m$ also a pure gluonic part. The aim of the Hamiltonian approach is to solve the Schr\"odinger equation $H \psi  [A] = E \psi [A]$ for the vacuum wave functional $\psi [A]$ in the Hilbert space with the scalar product
\beq
\langle \phi | \ldots | \psi \rangle = \int \cD A \, J[A] \phi^*[A] \ldots \psi[A] \, . \label{414-G4}
\eeq
Here the functional integration is over the transversal part of the spatial gauge field and the Fadeev--Popov determinant $J[A]$ arises by fixing to Coulomb gauge with the standard Fadeev--Popov method. To solve the Yang--Mills Schr\"odinger equation we use the variational principle with the following trial ansatz for the vacuum wave functional \cite{Feuchter2004, Feuchter2005} (for early attempts see refs.~\cite{Schuette1984, Szczepaniak2001})
\beq
\phi[A] = \frac{1}{\sqrt{J[A]}} \exp\left[-\frac{1}{2} \int \dd^3 x \int \dd^3 y \, A_k^a(\vx) \omega(\vx, \vy) A_k^a(\vy)\right] . \label{419-G5}
\eeq
This contains the variational kernel $\omega (\vx, \vy)$ which has to be determined by minimizing the vacuum energy $\langle H \rangle \equiv \langle \phi \vert H \vert \phi \rangle$. Due to translational and rotational invariance, the Fourier transform of the kernel is given by
\beq
\omega(\vx, \vy) = \int \frac{\dd^3 p}{(2 \pi)^3} \, \exp\bigl(\ii \vp \cdot (\vx - \vy)\bigr) \, \omega(p)
\eeq
where the kernel in momentum space has the meaning of the gluon energy and depends only on the modulus $p = |\vp|$. The result of the variational calculation is shown in fig.~\ref{fig1} (a). For large momenta, the gluon energy rises linearly with the modulus of the three-momentum in accordance with asymptotic freedom while in the infrared it diverges like $\omega(p) \sim 1/p$. Figure \ref{fig1} (b) confronts the variational result for the static gluon propagator \cite{ERS2007} with the lattice data \cite{BQR2009}. The agreement between the lattice and the continuum results is quite good in the infrared and also in the UV, however, the continuum result misses some strength in the mid-momentum regime. A better agreement with the lattice data is obtained by using a non-Gaussian wave functional \cite{CR2010}, where also cubic and quartic terms in the gauge fields were included in the exponent of the vacuum wave functional. This requires the use of Dyson--Schwinger equations to relate the variational kernel in the exponent of the wave functional to the $n$-point functions. For more details see D.~Campagnari's contribution to this conference.

\section{The Coulomb string tension}

\begin{figure}
\centering
\begin{subfigure}{0.45\textwidth}
\includegraphics[width=\textwidth,clip]{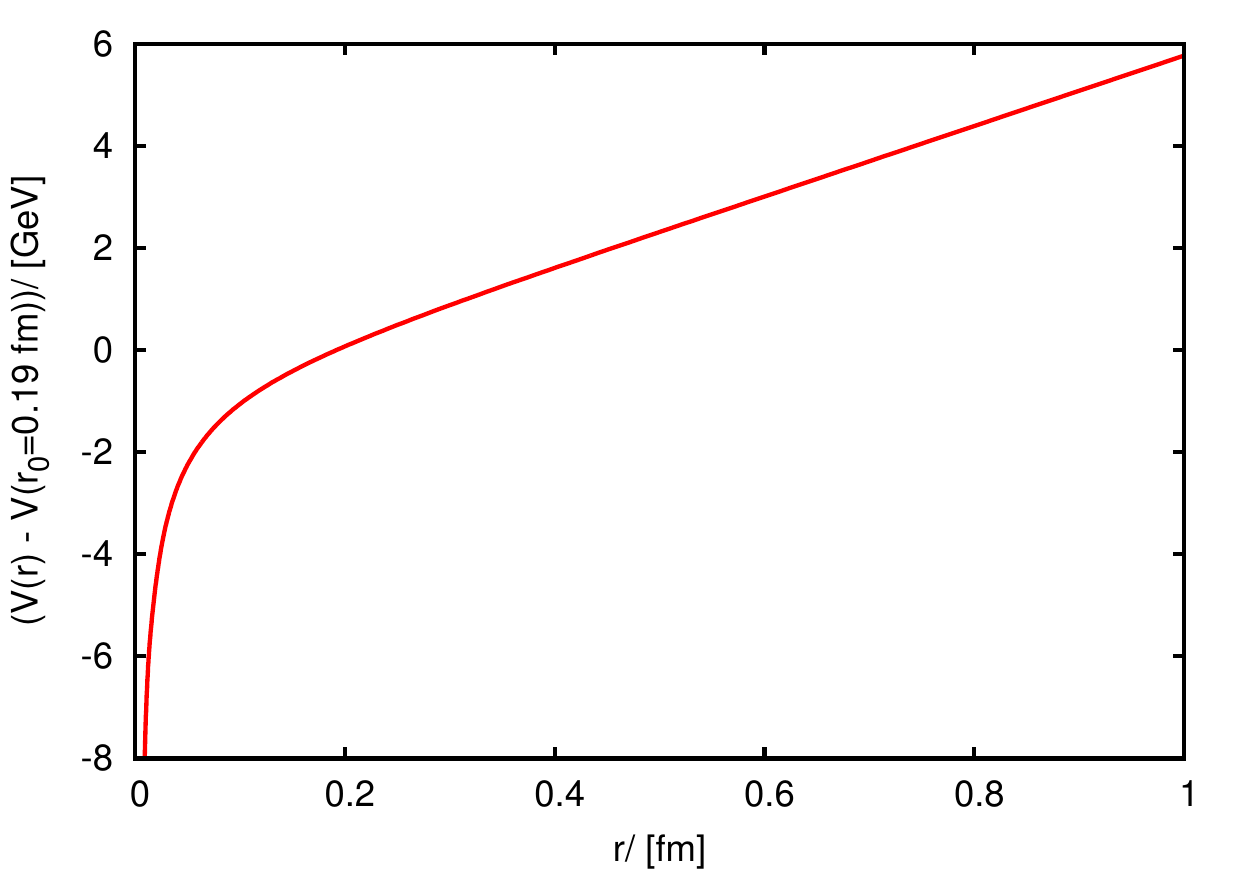}
\caption{}
\end{subfigure}
\quad
\begin{subfigure}{0.45\textwidth}
\includegraphics[width=\textwidth,clip]{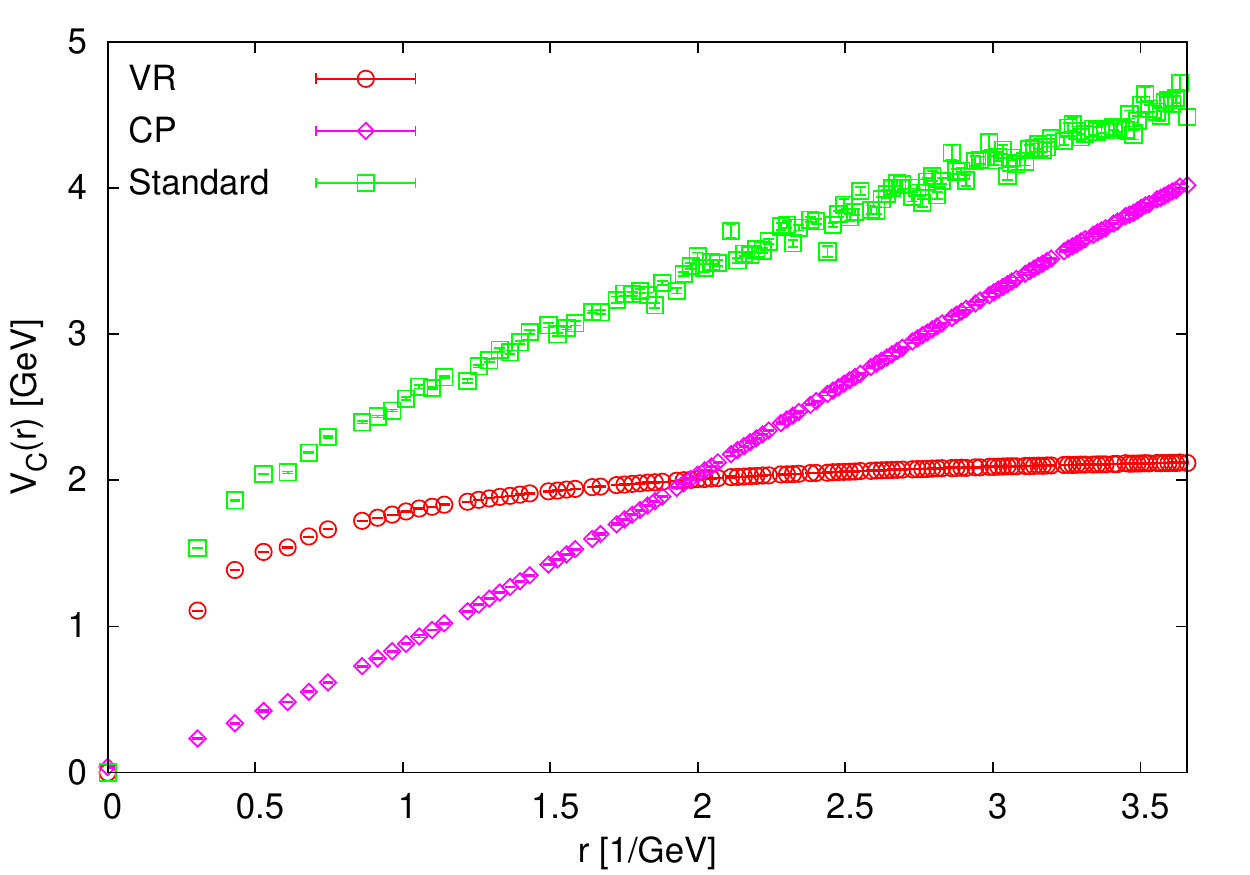}
\caption{}
\end{subfigure}
\caption{(a) Non-Abelian Coulomb potential obtained within the variational approach. (b) Standard non-Abelian Coulomb potential (green boxes) compared to the potential obtained after vortex removal (red circles) and center projection (violet diamonds).}
\label{fig2}%
\end{figure}

The Coulomb term $H_{\mathrm{C}}$ plays an important role in the Gribov--Zwanziger confinement scenario. Its Yang--Mills vacuum expectation value
\beq
V_{\mathrm{C}} = g^2 \bigl\langle (-\hat{\vD} \cdot \vec{\partial})^{-1} (-\vec{\partial}^2) (-\hat{\vD} \cdot \vec{\partial})^{-1} \bigr\rangle \label{Gx}
\eeq
provides an upper bound for the potential between static point-like color charges. The Coulomb potential found within the variational approach \cite{Feuchter2004, Feuchter2005} is shown in fig.~\ref{fig2} (a), \cite{ERS2007}. At small distances it behaves like an ordinary Coulomb potential, $V_{\mathrm{C}}(r) \sim 1/r$, and increases linearly at large distances with a coefficient given by the so-called Coulomb string tension $\sigma_{\mathrm{C}}$. It was shown in \cite{Zwanziger2003} that this quantity is an upper bound to the 
Wilsonian string tension $\sigma_{\mathrm{W}}$. On the lattice one finds $\sigma_{\mathrm{C}} / \sigma_{\mathrm W} \approx 2 \ldots 4$ \cite{BQRV2015, GOZ2004, Voigt2008}. In the Gribov--Zwanziger confinement scenario a necessary condition for confinement is that the non-Abelian Coulomb potential (\ref{Gx}) rises linearly at large distances due to the constraint $\sigma_{\mathrm{C}} \geq \sigma_{\mathrm{W}}$. For later consideration let us also mention that the Coulomb potential can be calculated on the lattice from the temporal 
links $U_0 (\vx)$ as \cite{Greensite:2003xf}
\be
\label{8}
a V_{\mathrm{C}}(| \vx - \vy|) = -\log\bigl\langle \tr \, \Bigl(U_0(\vx) U_0^{\dagger}(\vy)\Bigr) \bigr\rangle
\ee
where $a$ denotes the spacing of the lattice sites.

Lattice calculations provide strong evidence that confinement is due to extended gauge field configurations like center vortices. On the lattice center vortices can be detected as follows \cite{DelDebbio1998}: One first brings the gauge field configurations into the so-called maximal center gauge
\be
\label{9}
\sum_{x, \mu} \bigl\vert \tr U_{\mu}^2(x)\bigr\vert \to \max \, ,
\ee
which rotates a link as close as possible to a center element, $Z_\mu (y) = \pm 1 \in Z (2)$ for the gauge group SU(2). Subsequently, one performs a so-called center projection
\be
\label{10}
U_\mu (x) \to Z_\mu (x)
\ee
which replaces each link by its nearest center element. One is then left with $Z(2)$ links, which form closed center vortices, the only non-trivial field configurations in a $Z(2)$ theory. When a center vortex pierces a Wilson loop it contributes a non-trivial center element to the latter. It was shown in ref.~\cite{Langfeld1997} that the center vortices obtained in this way are physical objects in the sense that they show the proper scaling behavior, i.e.~their area density survives the continuum limit.

The center vortex content of a gauge field configuration can be removed \cite{deForcrand1999} by multiplying the original link variable $U_{\mu}(x)$ by its center projection $Z_{\mu}(x)$
\be
\label{11}
U_{\mu}(x) \to U_{\mu}(x) \cdot Z_{\mu} (x) \, .
\ee
One finds that after removal of center vortices the static color potential extracted from a Wilson loop loses its linearly rising part, e.g.~the Wilsonian string tension $\sigma_{\mathrm{W}}$ disappears after center vortex removal. Since $\sigma_{\mathrm{C}} \geq \sigma_{\mathrm{W}}$ this does not necessarily imply that elimination of center vortices also removes the Coulomb string tension. In ref.~\cite{BQRV2015} the non-Abelian Coulomb potential was calculated after center projection and center vortex removal. Removing the center vortices also eliminates the Coulomb string tension while center vortex projection keeps only the linearly rising part of the non-Abelian Coulomb potential. This result is perhaps not so surprising since center vortices live on the Gribov horizon\footnote{More precisely on the common boundary between the Gribov horizon and the fundamental modular region \cite{Greensite:2004ur}.}, which represents the domain of the infrared dominant field configuration in the Gribov--Zwanziger confinement scenario.

\begin{figure}
\centering
\begin{subfigure}{0.45\textwidth}
\includegraphics[width=\textwidth,clip]{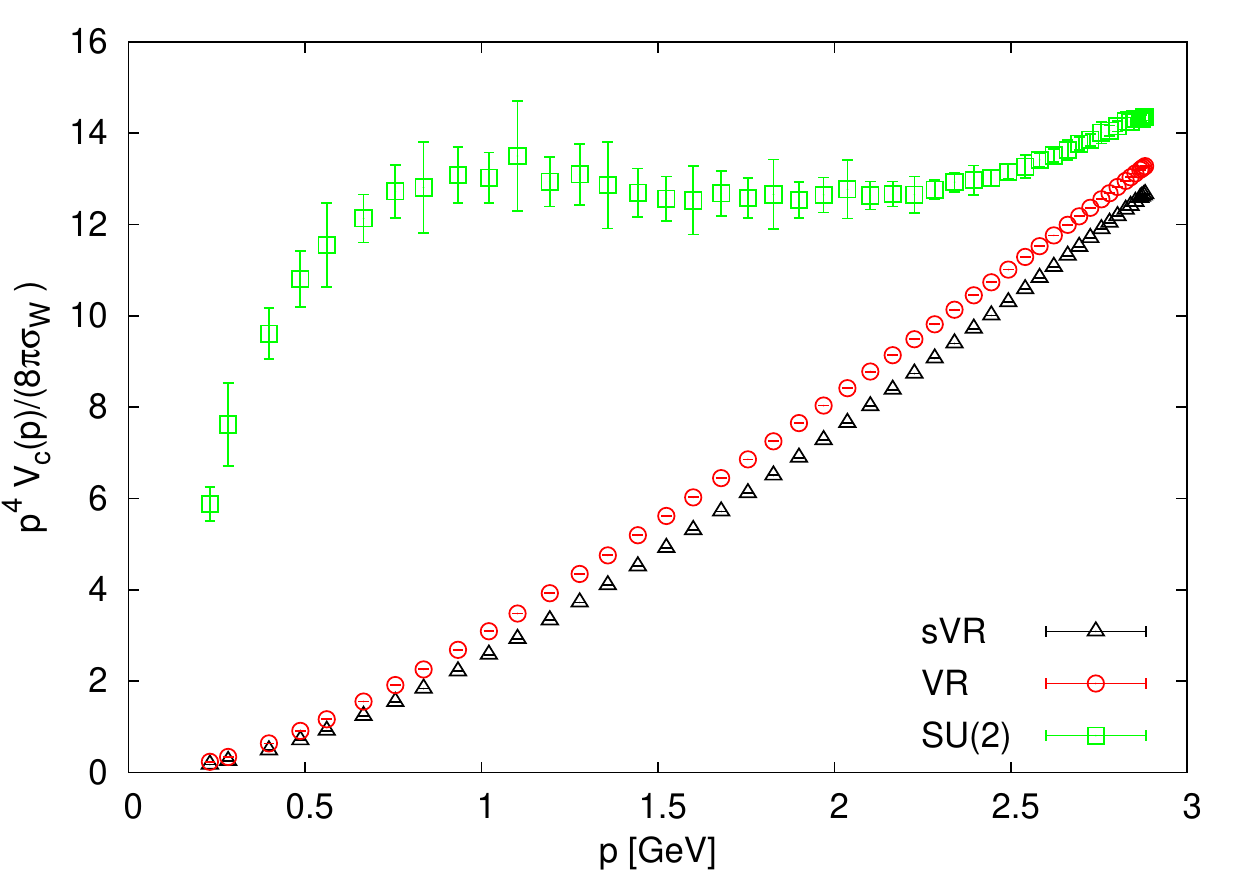}
\caption{}
\end{subfigure}
\quad
\begin{subfigure}{0.45\textwidth}
\includegraphics[width=\textwidth,clip]{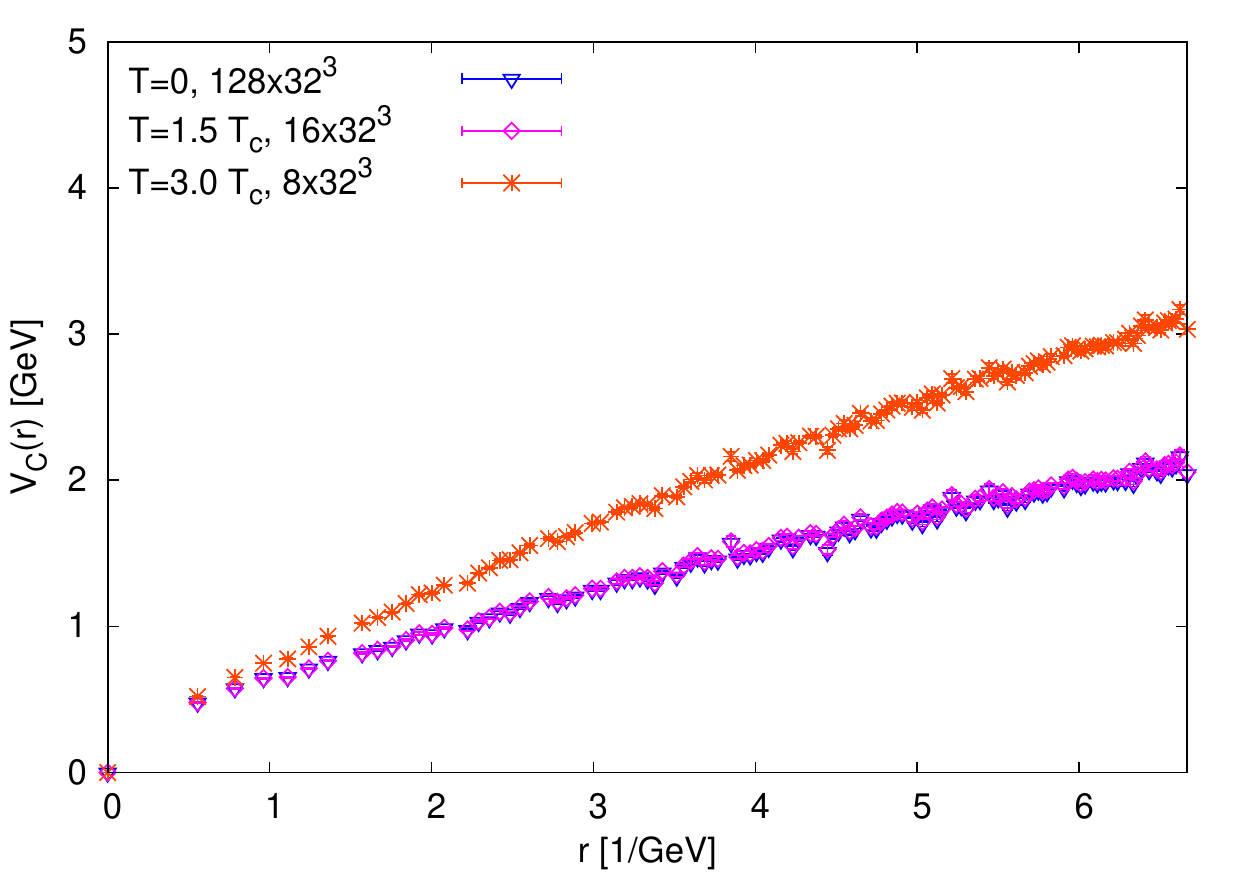}
\caption{}
\end{subfigure}
\caption{(a) Non-Abelian Coulomb potential in momentum space (green boxes) compared to the result obtained after removing just the spatial (black triangles) or all center vortices (red circles). (b) Non-Abelian Coulomb potential for different temperatures ($T_{\mathrm{c}}$ denotes the critical temperature).}
\label{fig3}%
\end{figure}

At finite temperature different Wilsonian string tensions are measured from temporal and spatial Wilson loops referred to as temporal and spatial string tension, respectively. Above the deconfinement phase transition these two Wilsonian string tensions decouple. While the spatial string tension increases above the critical temperature, the temporal string tension disappears. On the lattice it is not difficult to see that in the center projected $Z(2)$ theory the temporal and spatial Wilsonian string tension, i.e.~the area law in the temporal and spatial Wilson loop, are produced by temporal and spatial center vortices, respectively. The latter are formed exclusively by spatial center elements and are obtained by performing the center projection only for the spatial links
\be
\label{12}
U_i (x) \to Z_i (x) \, ,
\ee
which will be referred to as spatial center projection in the following. Analogously multiplying the spatial link by its nearest center projected $Z(2)$ element
\be
\label{13}
U_i(x) \to U_i(x) \cdot Z_i(x)
\ee
removes all spatial center vortices and thus the spatial string tension while the temporal links are unaffected. The temporal string tension, which can be exclusively calculated from the temporal links, will not be affected by the spatial center vortex removal. Figure \ref{fig3} (a) shows the quantity $p^4 V_{\mathrm{C}}(p)$ whose infrared limit gives the Coulomb string tension, $\lim_{p \to 0} p^4 V_{\mathrm{C}}(p) = 8 \pi \sigma_{\mathrm{C}}$. As one observes the Coulomb string tension disappears already when only the spatial center vortices are removed. This clearly shows that the Coulomb string tension is related to the spatial string tension and not to the temporal one. This explains also the finite-temperature behavior of the Coulomb string tension, which increases with the temperature above the deconfinement phase transition just like the spatial string tension, see fig.~\ref{fig3} (b).

\begin{figure}
\centering
\begin{subfigure}{0.45\textwidth}
\includegraphics[width=\textwidth,clip]{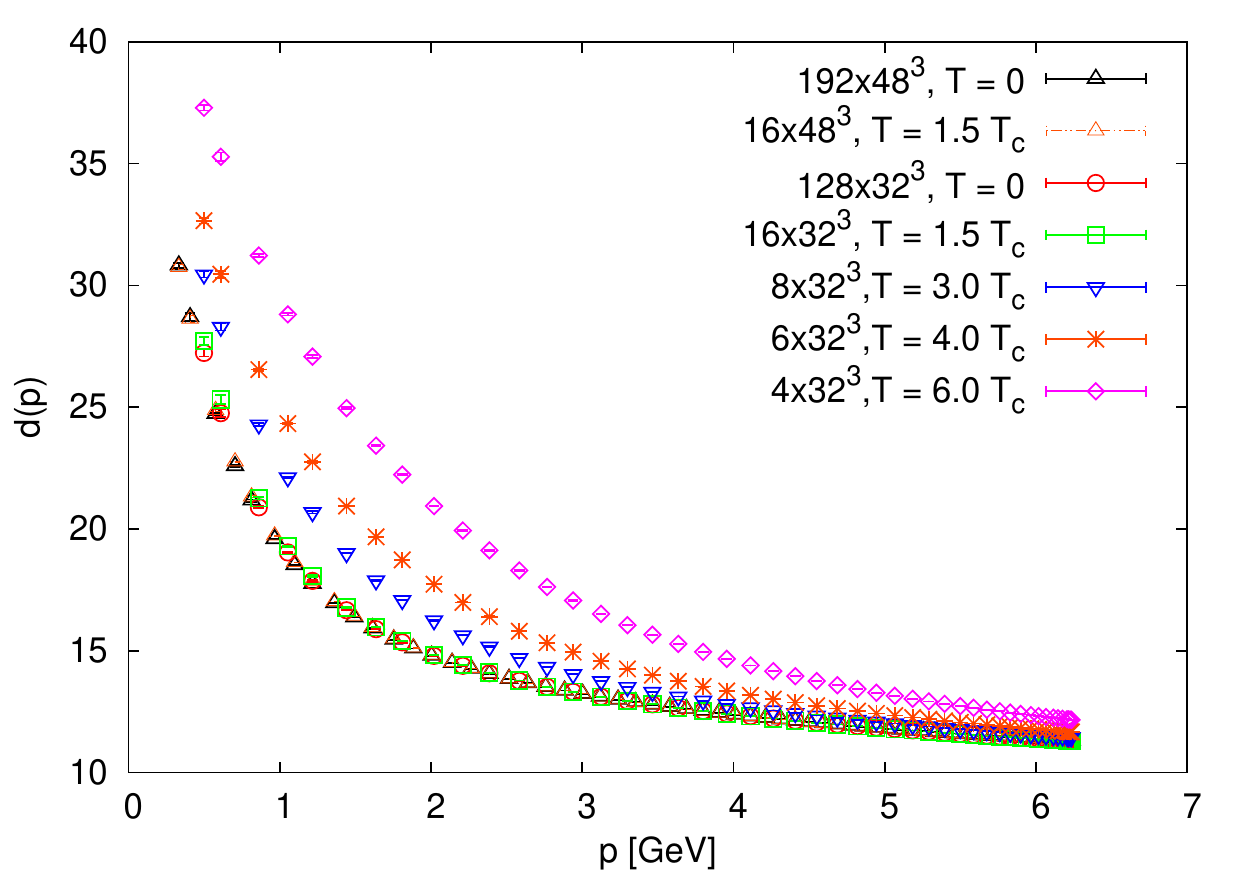}
\caption{}
\end{subfigure}
\quad
\begin{subfigure}{0.45\textwidth}
\includegraphics[width=\textwidth,clip]{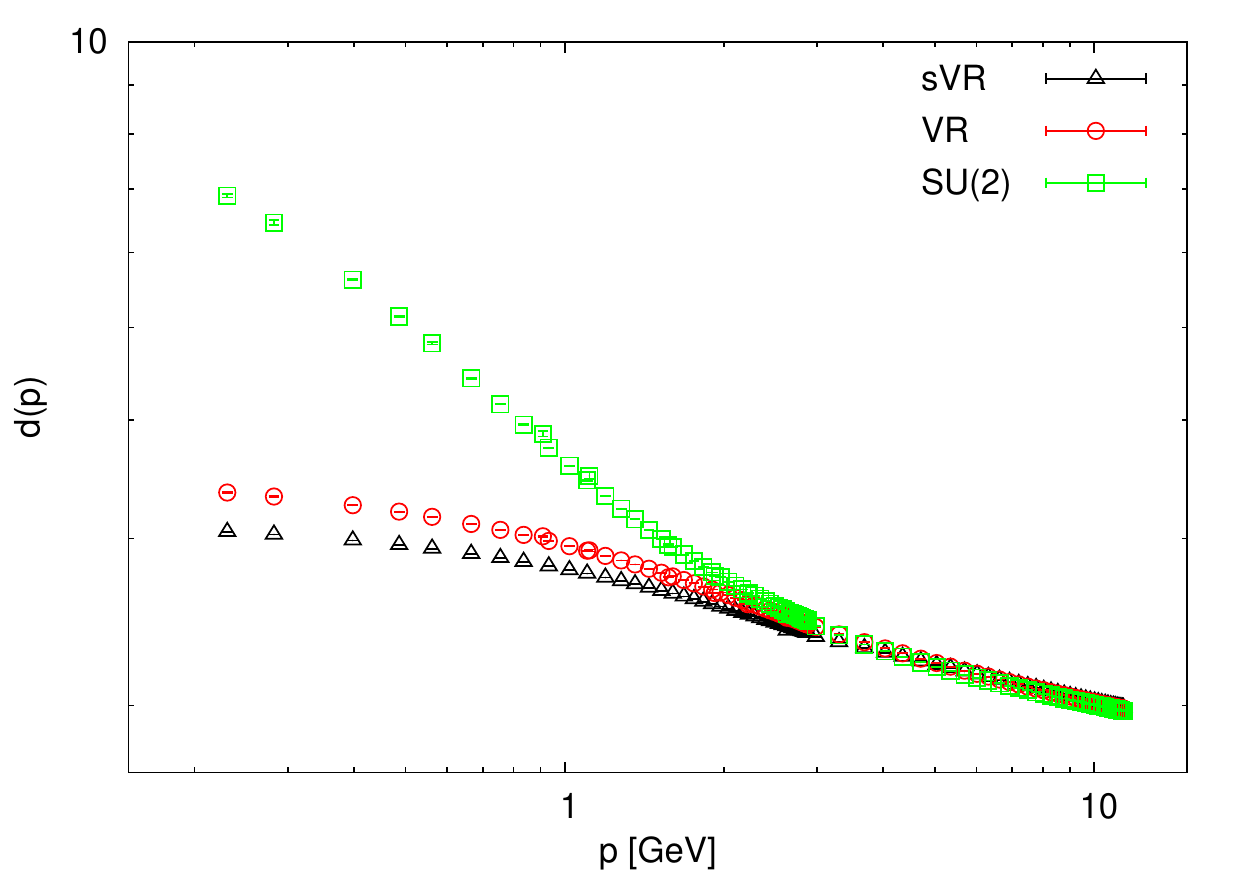}
\caption{}
\end{subfigure}
\caption{(a) Ghost form factor calculated on the lattice for different temperatures. (b) Ghost form factor (green boxes) compared to the result obtained after removing the spatial (black triangles) or all center vortices (red circles).}
\label{fig7}%
\end{figure}

A necessary condition for the Gribov--Zwanziger confinement scenario to be realized is that the ghost form factor is infrared divergent, which is indeed found in the variational approach and also on the lattice, see fig.~\ref{fig7} (a). However, the infrared divergence disappears when one removes the center vortices or the spatial center vortices only as can be seen in fig.~\ref{fig7} (b). Also, the spatial center vortex projection produces the same ghost form factor as full center projection. This also explains why the infrared divergence of the ghost form factor does not disappear above the deconfinement phase transition. Thus both features of the Gribov--Zwanziger confinement scenario, the infrared diverging ghost form factor and the linearly rising Coulomb potential, are caused by spatial center vortices and are thus tied to the spatial string tension.

\section{Variational approach to the quark sector}

The variational approach to Yang--Mills theory in Coulomb gauge presented in section \ref{sectII} has been extended to full QCD in refs.~\cite{Pak2013, QCDT0, Campagnari:2016wlt}. The Hamiltonian of full QCD in Coulomb gauge is given by
\be
\label{14}
H_{\mathrm{QCD}} = H_{\mathrm{T}} + H_{\mathrm{Q}} + H_{\mathrm{C}} \, ,
\ee
where $H_{\mathrm{T}}$ is the Hamiltonian of the transversal gluon degrees of freedom (\ref{G1}), $H_{\mathrm{C}}$ is the Coulomb interaction (\ref{2}) and
\beq
H_{\mathrm{Q}} = \int \dd^3 x \, \psi^\dagger(\vx) \Bigl[\vec{\alpha} \cdot \bigl(-\ii \nabla + g t^a \vA^a(\vx)\bigr) + \beta m_0\Bigr] \psi(\vx) \label{15}
\eeq
is the Hamiltonian of the quarks coupling to the transversal gluon field. Here, $\vec{\alpha}$, $\beta$ are the usual Dirac matrices, $t^a$ denotes the generator of the color group in the fundamental representation and $m_0$ is the bare (electroweak) current quark mass which will be neglected in the following. Furthermore, when the quarks are included, the matter charge density in the Coulomb Hamiltonian $H_{\mathrm{C}}$ (\ref{3}) is given by
\be
\label{16}
\rho_m^a(\vx) = \psi^\dagger(\vx) t^a \psi(\vx) \, .
\ee
In refs.~\cite{QCDT0, Campagnari:2016wlt}, the quark sector of QCD has been treated within the variational approach using the following ansatz for the QCD wave functional
\beq
\vert \phi[A] \rangle = \phi_{\mathrm{YM}}[A] \, \vert \phi_{\mathrm{Q}}[A] \rangle \label{18}
\eeq
where $\phi_{\mathrm{YM}}$ is essentially given by the Yang--Mills vacuum functional (\ref{419-G5}) and
\beq
\vert \phi_{\mathrm{Q}}[A] \rangle = \exp\left[-\int \dd^3 x \int \dd^3 y \, \psi_+^{\dagger}(\vx) K(\vx, \vy) \psi_-(\vy)\right] \vert 0 \rangle \, , \label{17}
\eeq
with
\beq
K(\vx, \vy) = \beta S(\vx, \vy) + g \int \dd^3 z \, \bigl[V(\vx, \vy; \vz) + \beta W(\vx, \vy; \vz)\bigr] \vec{\alpha} \cdot \vA^a(\vz) t^a \label{ansx}
\eeq
is the quark wave functional. Here $S$, $V$ and $W$ are variational kernels. Furthermore, $\vert 0 \rangle$ is the Fock vacuum of the quarks which represents the bare Dirac sea.

The ansatz (\ref{17}) reduces for $W = 0$ to the quark wave functional used in ref.~\cite{Pak2013} while for $V = W = 0$ it becomes the BCS-type wave functional considered in refs.~\cite{FM1982, Adler1984, AA1988}. With the wave functional (\ref{18}) the expectation value of the QCD Hamiltonian was calculated up to two loops. Variation with respect to the two kernels $V$ and $W$, which describe the coupling of the quarks to the transversal gluons, gives two equations, which can be solved explicitly in terms of the scalar kernel $S$ and the gluon energy $\omega$ yielding
\begin{align}
V(\vp, \vq) &= \frac{1 + S(p) S(q)}{p P(p) \Bigl(1 - S^2(p) + 2 S(p) S(q)\Bigr) + q P(q) \Bigl(1 - S^2(q) + 2 S(p) S(q)\Bigr) + \omega(|\vp + \vq|)} \, , \label{19} \\
W(\vp, \vq) &= \frac{S(p) + S(q)}{p P(p) \Bigl(1 - S^2(p) - 2 S(p) S(q)\Bigr) + q P(q) \Bigl(1 - S^2(q) - 2 S(p) S(q)\Bigr) + \omega(|\vp + \vq|)} \label{20}
\end{align}
where we have defined the quantity
\be
P(p) = \frac{1}{1 + S^2(p)} \, . \label{23}
\ee
The variational equation for the scalar kernel $S$, referred to as \textit{gap equation}, is highly non-local and can only be solved numerically. However, one can show analytically that all UV divergences in this equation cancel: the UV-divergent contributions induced by the kernels $V$ and $W$ are given, respectively, by
\begin{align}
&\frac{C_{\mathrm{F}}}{16 \pi^2} g^2 S(k) \left[-2 \Lambda + k \ln \frac{\Lambda}{\mu} \left(-\frac{2}{3} + 4 P(k)\right)\right] , \label{21} \\
&\frac{C_{\mathrm{F}}}{16 \pi^2} g^2 S(k) \left[2 \Lambda + k \ln \frac{\Lambda}{\mu} \left(\frac{10}{3} - 4 P(k)\right)\right] . \label{22}
\end{align}
Here, $C_{\mathrm{F}} = (N_{\mathrm{C}}^2 - 1) / 2 N_{\mathrm{C}}$ is the quadratic Casimir, $\Lambda$ is the UV cutoff and $\mu$ is an arbitrary momentum scale. In the sum of the two terms given by eq.~(\ref{21}) and (\ref{22}) the linear UV divergences obviously cancel. Furthermore, the sum of the logarithmic UV divergences of these two terms cancel against the asymptotic contribution to the gap equation induced by the Coulomb kernel,
\beq
-\frac{C_{\mathrm{F}}}{6 \pi^2} g^2 k S(k) \ln \frac{\Lambda}{\mu} \, . \label{24}
\eeq

\begin{figure}
\centering
\begin{subfigure}{0.45\textwidth}
\includegraphics[width=\textwidth,clip]{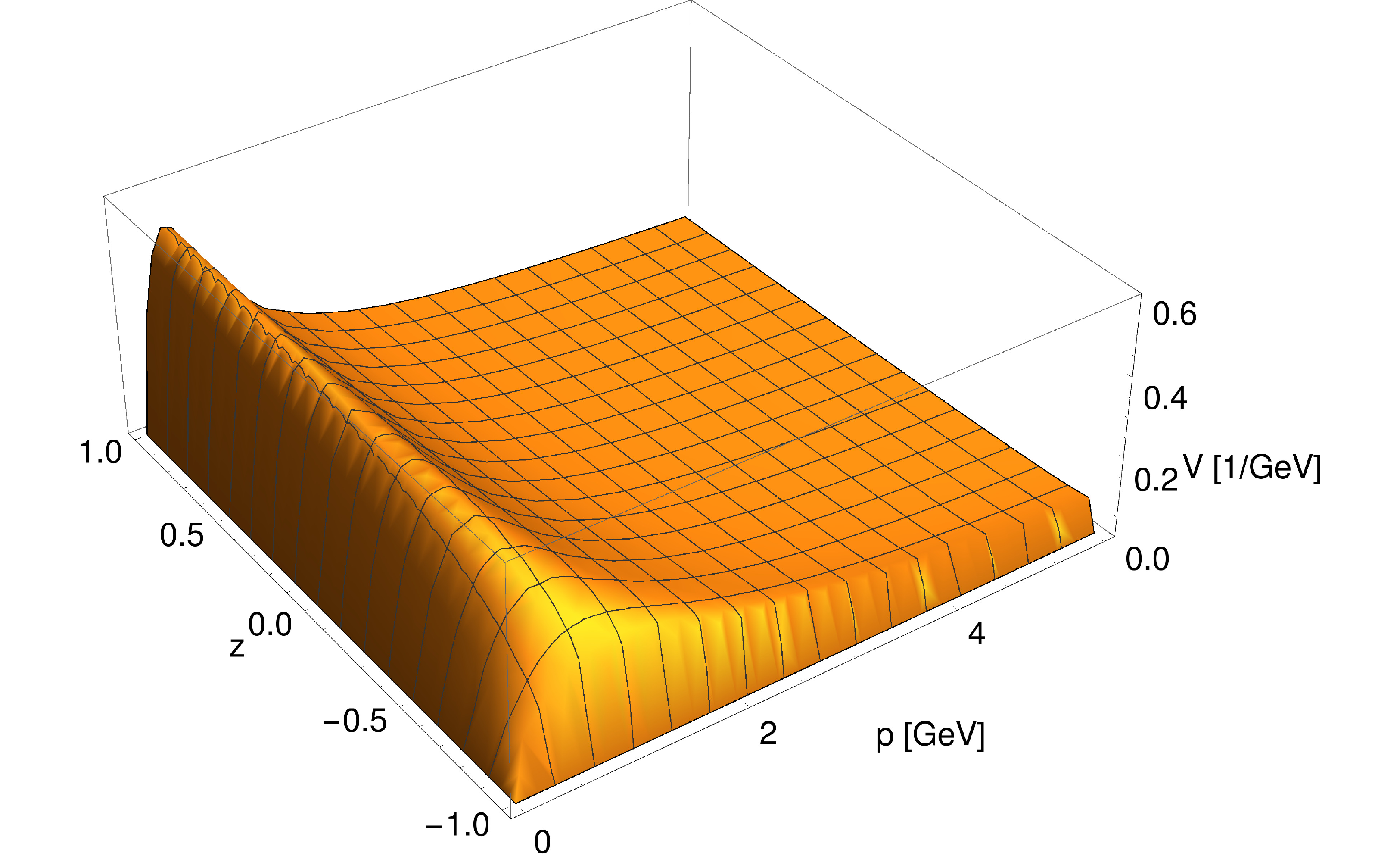}
\caption{}
\end{subfigure}
\quad
\begin{subfigure}{0.45\textwidth}
\includegraphics[width=\textwidth,clip]{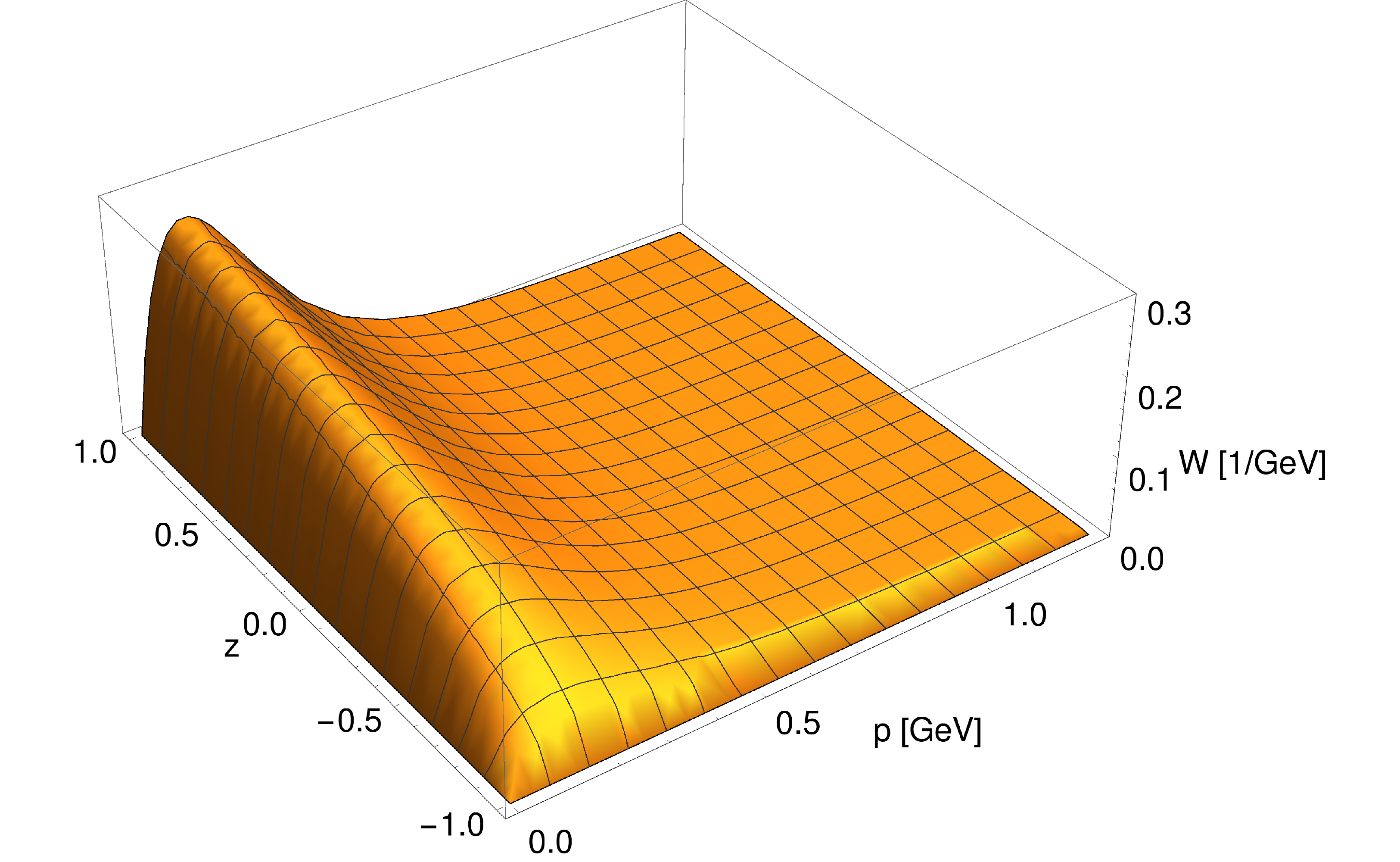}
\caption{}
\end{subfigure}
\caption{The vector kernel (a) $V(\vp, \vq)$ (\ref{19}) and (b) $W(\vp, \vq)$ (\ref{20}) obtained from the solution of the quark gap equation for $g \simeq 2.1$ as function of $p = q$ and $z = \cos\sphericalangle(\vp,\vq)$ \cite{Campagnari:2016wlt}.}
\label{fig4}%
\end{figure}%

\begin{figure}
\centering
\begin{subfigure}{0.45\textwidth}
\includegraphics[width=\textwidth,clip]{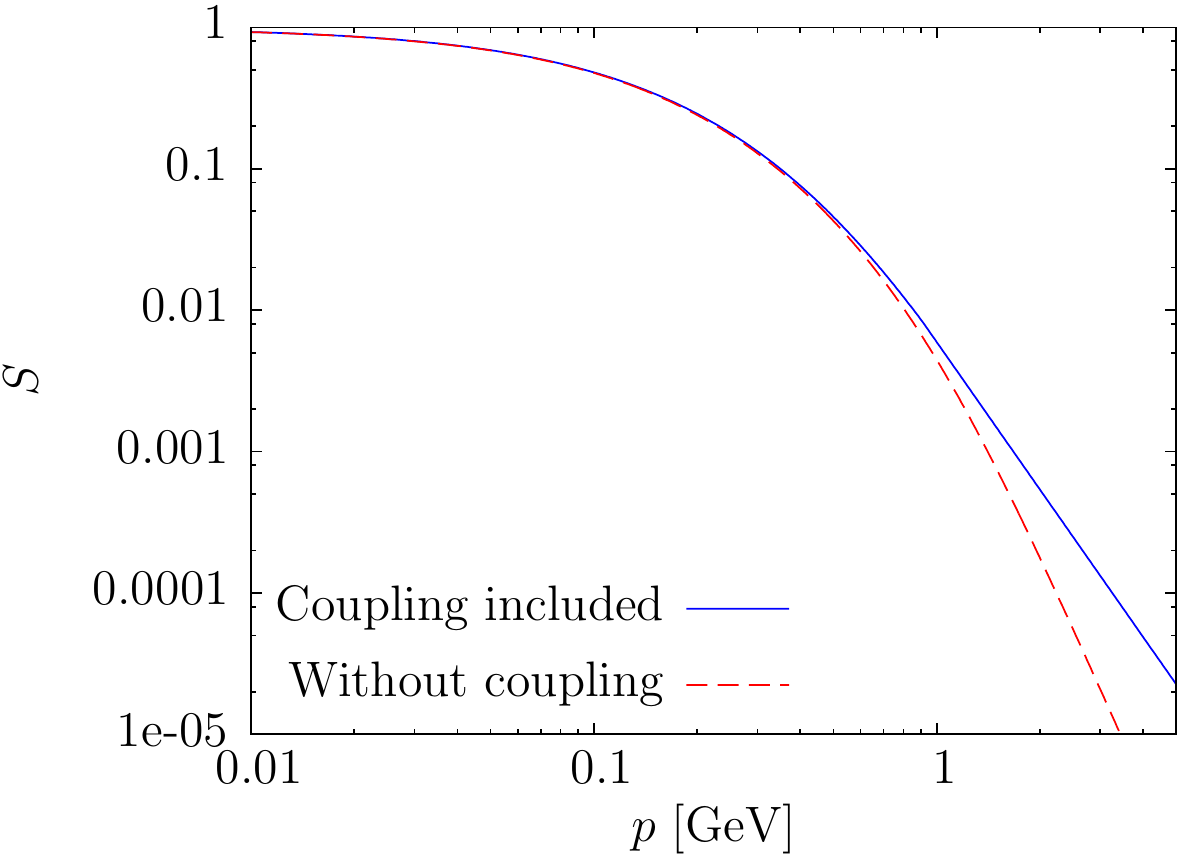}
\caption{}
\end{subfigure}
\quad
\begin{subfigure}{0.45\textwidth}
\includegraphics[width=\textwidth,clip]{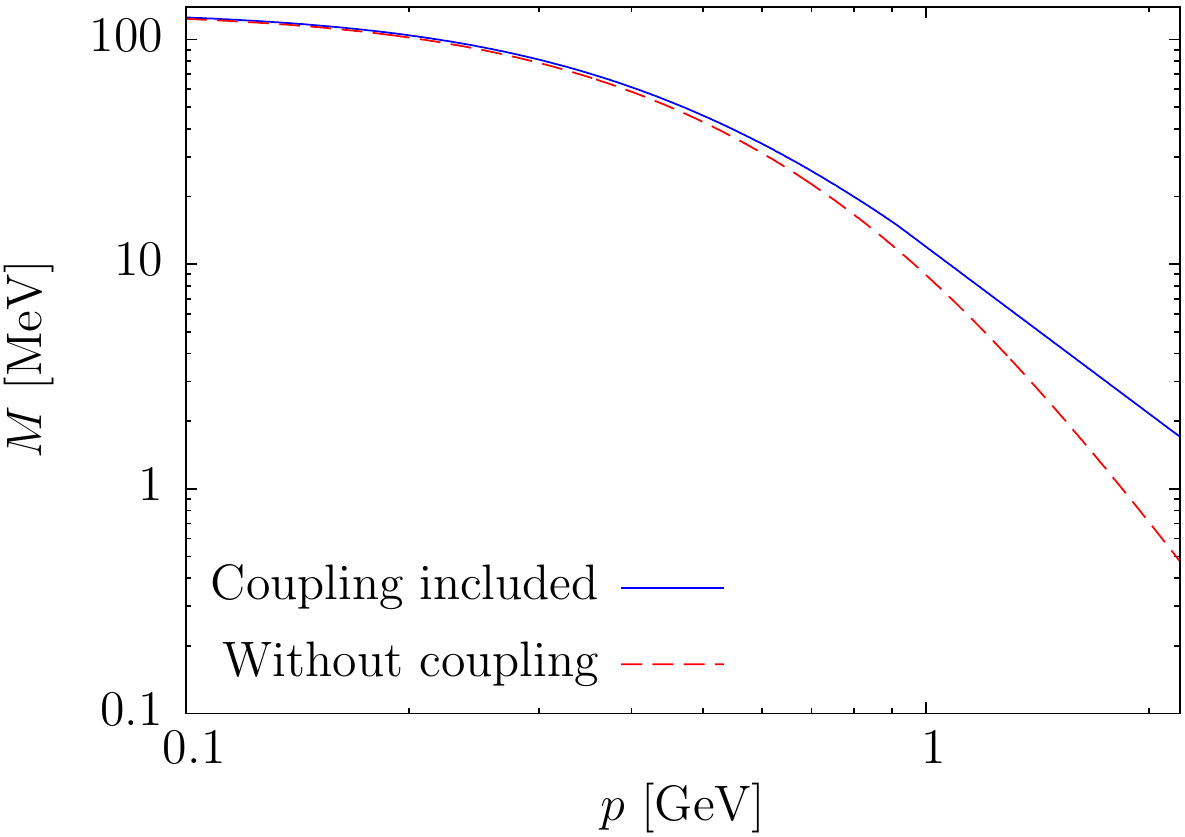}
\caption{}
\end{subfigure}
\caption{(a) Scalar form factor and (b) mass function obtained from the (quenched) solution of the quark gap equation. Results are presented for $g \simeq 2.1$ (full curve) and $g = 0$ (dashed curve).}
\label{fig5}%
\end{figure}%

Due to the exact cancellation of all UV divergences no renormalization of the gap equation is required. This is certainly a big advantage of the present ansatz (\ref{17}) for the quark wave functional. Using the gluon propagator $\sim 1/\omega$ obtained in the Yang--Mills sector as input, the quark gap equation can be solved within a quenched calculation. In this approach, the coupling constant $g$ is determined by fixing the chiral condensate to its phenomenological value \cite{Campagnari:2016wlt}. Figure \ref{fig4} shows the vector kernels $V(\vp, \vq)$, $W(\vp, \vq)$ obtained in this way, as function of the modulus $p = q$ of the ingoing quark momenta and the cosine of the angle between them, $z = \cos\sphericalangle(\vp,\vq)$. These kernels are peaked in the mid-momentum regime. Furthermore, the vector kernel $V$ is about a factor of two larger than the kernel $W$. Figure \ref{fig5} shows the scalar kernel $S(p)$ and the mass function
\be
M(p) = \frac{2 p S(p)}{1 - S^2(p)} \label{25}
\ee
on a logarithmic scale. For sake of comparison we also quote the curves obtained when the coupling to the transversal gluons is neglected. More precisely, this corresponds to putting $g = 0$ in the ansatz  (\ref{ansx}) and discarding the second (perturbative) part in the approximation
\be
V_{\mathrm{C}}(p)\approx \frac{8 \pi \sigma_{\mathrm{C}}}{p^4} + \frac{g^2}{p^2} \label{foo}
\ee
for the Coulomb potential eq.~(\ref{Gx}). As one observes the inclusion of the coupling to the transversal gluon changes only the mid- and large-momentum regime while the infrared behavior is not changed at all. This is perhaps a little bit surprising but should have been expected in view of the fact that the non-Abelian Coulomb term (the first part in eq.~(\ref{foo})) which gives rise to a linearly rising potential at large distances dominates the infrared behavior of the gap equation. Let us also mention that we do not find chiral symmetry breaking from our equations when the linearly rising part of the Coulomb potential is neglected.

\section{Hamiltonian approach to finite temperature QCD by compactifying a spatial dimension}

In refs.~\cite{Reinhardt2011, Heffner2012} the variational approach to Yang--Mills theory in Coulomb gauge was extended to finite temperatures by making a quasi-particle ansatz for the density matrix of the grand canonical ensemble where the quasi-particle energy was determined by minimizing the free energy. The resulting variational equations could be solved analogously to the ones at zero temperature. There is, however, a more efficient way to treat Yang--Mills theory at finite temperature within the Hamiltonian approach. The motivation comes from the Polyakov loop
\beq
P[A_0](\vx) = \frac{1}{d_r} \tr P \exp\left[\ii \il_0^{\beta} \dd x^4 \, A_0 (\vx, x^4) \right] \, ,
\eeq
where $P$ is the path ordering prescription, $d_r$ denotes the dimension of the representation of the gauge group and $\beta$ is the length of the compactified Euclidean time axis which represents the inverse temperature. This quantity cannot be calculated straightforwardly in the Hamiltonian approach due to the unrestricted time interval and the use of the Weyl gauge $A_0 = 0$. An alternative, more efficient Hamiltonian approach to finite-temperature quantum field theory has been proposed in ref.~\cite{Reinhardt:2016xci}. It does not require an ansatz for the density matrix of the grand canonical ensemble and allows the evaluation of the Polyakov loop. In this novel approach one exploits the $O(4)$ invariance to interchange the Euclidean time axis with one spatial axis. The temporal (anti-)periodic boundary conditions to the fields become then spatial boundary conditions, while the new (Euclidean) time axis has infinite extent as is required within the Hamiltonian approach. The upshot is that the partition function at finite temperature $\beta^{- 1}$ is entirely given by the ground state calculated on the spatial manifold $\R^2 \times S^1(\beta)$. The whole thermodynamics of the theory is then encoded in the vacuum calculated on the partially compactified spatial manifold $\R^2 \times S^1(\beta)$. This approach was used in ref.~\cite{Heffner2015} to study Yang--Mills theory at finite temperature and in ref.~\cite{Reinhardt:2013iia} to calculate the Polyakov loop within the Hamiltonian approach. In ref.~\cite{Reinhardt:2016pfe} this approach was used to calculate the so-called dual quark condensate.

The dual quark condensate was originally introduced in ref.~\cite{Gattringer2006} and was discussed in a more general context in ref.~\cite{Synatschke:2007bz}. This quantity has been calculated on the lattice \cite{Bilgici:2008qy} in the functional renormalization group approach \cite{Braun:2009gm} and in the Dyson--Schwinger approach \cite{FMM2010}. The dual condensate is defined by
\beq
\Sigma_n = \int\limits_0^{2 \pi} \frac{\dd \varphi}{2 \pi} \exp(-\ii n \varphi) \langle \bar{\psi} \psi \rangle_\varphi \, , \label{27}
\eeq
where $\langle \bar{\psi} \psi \rangle_\varphi$ is the quark condensate calculated with the $U(1)$-valued boundary condition
\beq
\psi(x_4 + \beta/2, \vx) = \mathrm{e}^{\ii \varphi} \psi(x_4 - \beta/2, \vx) \, . \label{28}
\eeq
For $\varphi = \pi$ these boundary conditions reduce to the usual finite-temperature boundary conditions of the quark field in the functional integral representation of the partition function. On the lattice it is not difficult to show that the quantity $\Sigma_n$ (\ref{27}) represents the vacuum expectation value of the sum of all closed Wilson loops winding precisely $n$-times around the compactified time axis. In particular, the quantity $\Sigma_1$ represents the expectation value of all closed loops winding precisely once around the compactified time axis and is therefore called the dressed Polyakov loop. The phase in the boundary condition (\ref{28}) can be absorbed into an imaginary chemical potential
\beq
\mu = \ii \frac{\pi - \varphi}{\beta} \label{29}
\eeq
for fermion fields satisfying the usual antisymmetric boundary condition $\psi(x_4 + \beta/2, \vx) = -\psi(x_4 - \beta/2, \vx)$. In the Hamiltonian approach to finite temperatures of ref.~\cite{Reinhardt:2016xci} where the compactified time axis has become the third spatial axis the phase dependent boundary condition (\ref{28}) or equivalently the imaginary chemical potential (\ref{29}) manifests itself in the momentum variable along the (compactified) 3-axis, which reads
\beq
p_3 = \Omega_n + \ii \mu = \frac{2 \pi n + \varphi}{\beta} \, , \quad \quad \Omega_n = \frac{(2 n + 1) \pi}{\beta} \, , \label{30}
\eeq
where $\Omega_n$ is the usual fermionic Matsubara frequency. Using the zero temperature quark mass function $M(p)$ calculated in ref.~\cite{Campagnari:2016wlt}, one finds in the Hamiltonian approach to QCD of ref.~\cite{QCDT0} for the dual quark condensate after Poisson resummation the leading expression \cite{Reinhardt:2016pfe}
\beq
\Sigma_n = -\frac{N_{\mathrm{C}}}{\pi^2} \int\limits_0^{\infty} \dd p \, \frac{p^2 M(p)}{\sqrt{p^2 + M^2(p)}} \left[\delta_{n 0} + \frac{\sin(n \beta p)}{n \beta p}\right] \, . \label{31}
\eeq
\begin{figure}
\centering
\begin{subfigure}{0.45\textwidth}
\includegraphics[width=\textwidth,clip]{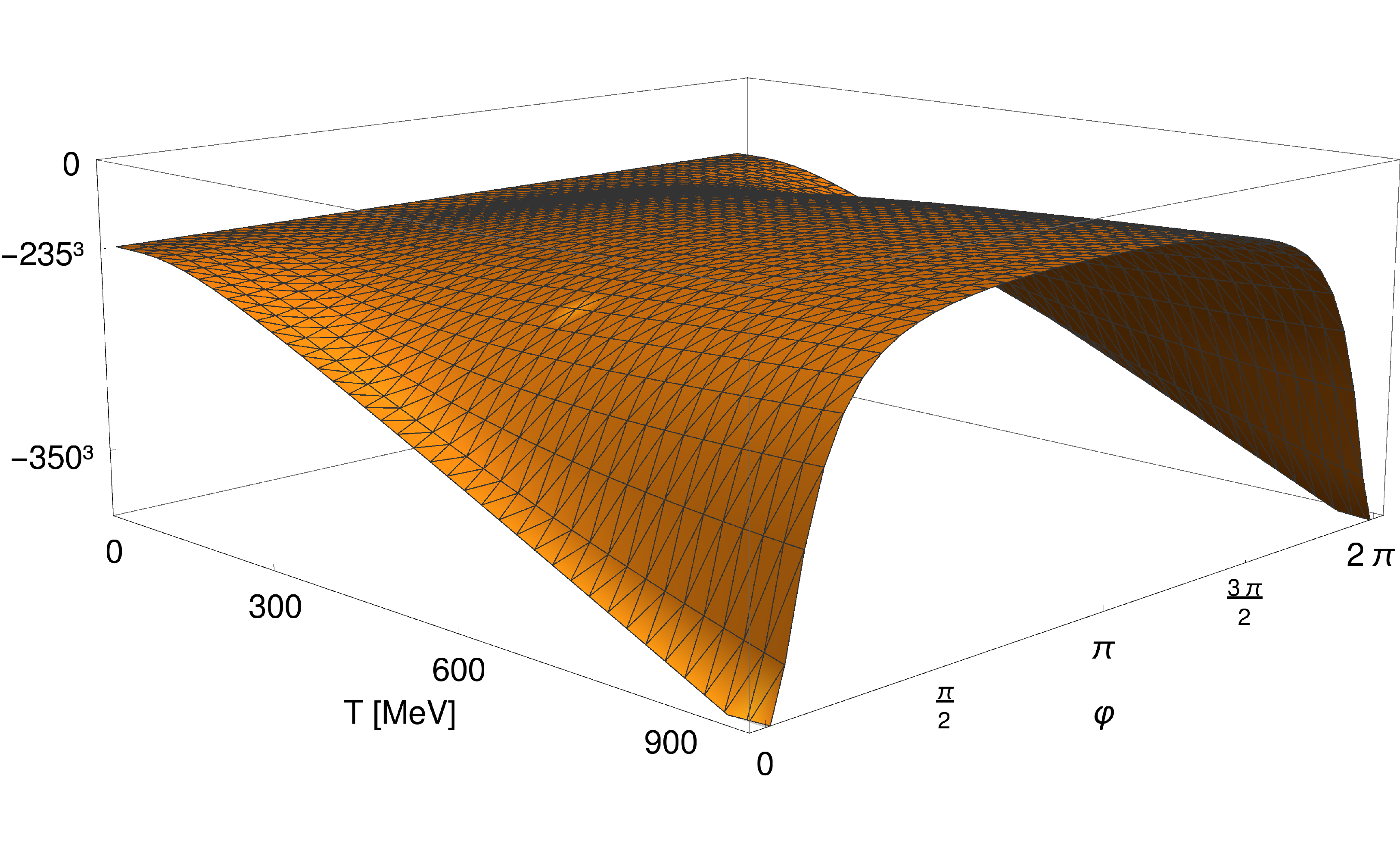}
\caption{}
\end{subfigure}
\quad
\begin{subfigure}{0.45\textwidth}
\includegraphics[width=\textwidth,clip]{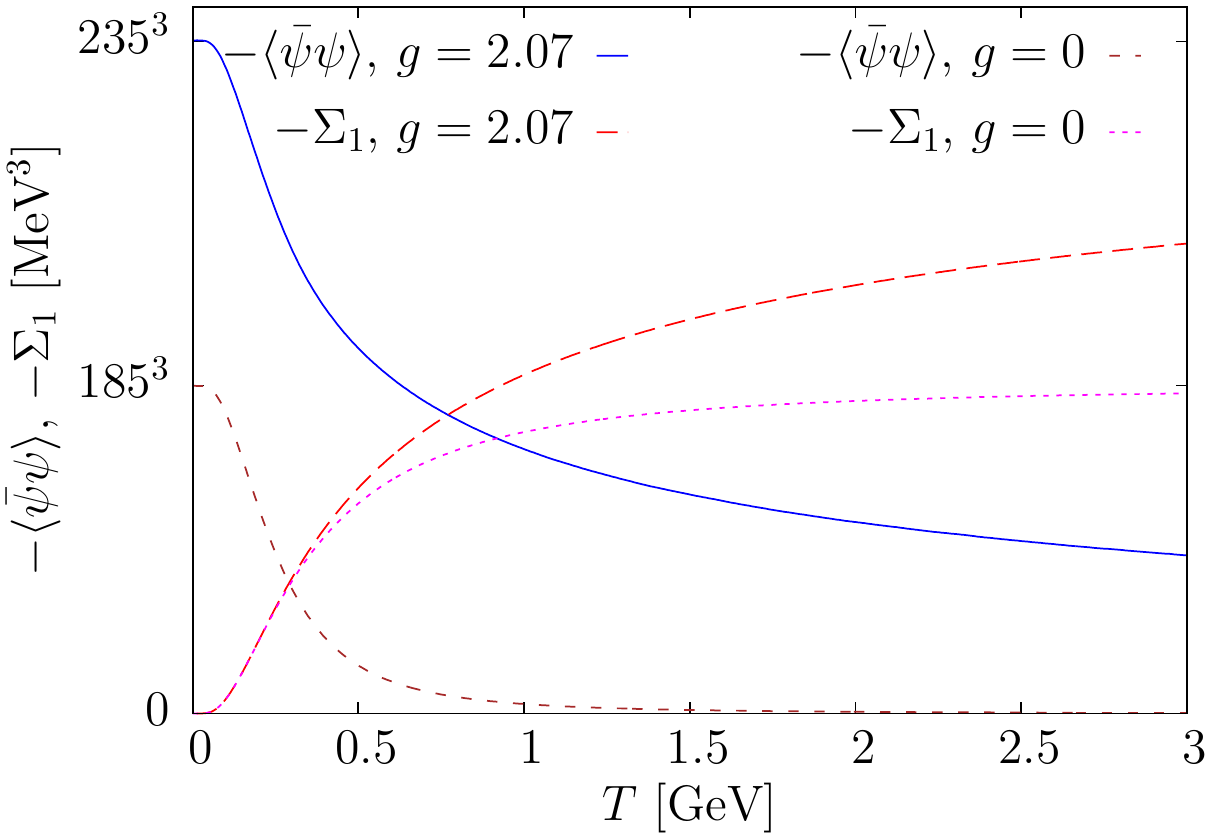}
\caption{}
\end{subfigure}
\caption{(a) Chiral quark condensate $\langle \bar{\psi} \psi \rangle_\varphi$ as function of the temperature $T$ and the phase $\varphi$ of the boundary condition (\ref{28}). (b) Chiral and dual quark condensate as function of the temperature. Results are presented for both a coupling of $g \simeq 2.1$ and $g = 0$.}
\label{fig6}%
\end{figure}%
In the same line, one finds the quark condensate $\langle \bar{\psi} \psi \rangle_\varphi$ shown in fig.~\ref{fig6} (a). For the dressed Polyakov loop one finds the temperature behavior shown in fig.~\ref{fig6} (b), where we also compare with the result obtained when the coupling to the transversal gauge field degrees of freedom is neglected ($g = 0$). As one observes there is no difference at small temperatures in accord with the fact that the mass function $M(p)$ has the same infrared behavior, whether the coupling to the transversal gluons is included or not. The slower UV decrease of the full mass function causes the dual condensate to reach its high-temperature limit
\beq
\lim_{\beta \to 0} \Sigma_1 = -\frac{N_{\mathrm{C}}}{\pi^2} \int\limits_0^{\infty} \dd p \, \frac{p^2 M(p)}{\sqrt{p^2 + M^2(p)}} = \lim_{\beta \to \infty} \langle \bar{\psi} \psi \rangle_{\varphi = \pi} \label{32}
\eeq
only very slowly. We expect, however, that this limit is reached faster when the finite-temperature solutions are used. This will presumably also convert the crossover obtained for the chiral condensate, see fig.~\ref{fig6} (b), into a true phase transition as expected for chiral quarks. From the inflexion points of the chiral and dual condensate one extracts the values of $T_{\chi}^{\mathrm{pc}} \simeq 170 \, \mathrm{MeV}$ and $T_{\mathrm{C}}^{\mathrm{pc}} \simeq 198 \, \mathrm{MeV}$ for the pseudo-critical temperatures of the chiral and deconfinement transition, respectively. On the lattice one finds for realistic quark masses $T_{\chi}^{\mathrm{pc}} \simeq 155 \, \mathrm{MeV}$ and $T_{\mathrm{C}}^{\mathrm{pc}} \simeq 165 \, \mathrm{MeV}$ \cite{Borsanyi2010, Bazavov2012}.

\section{Conclusions}

In my talk I have presented some recent results obtained within the Hamiltonian approach to QCD in Coulomb gauge. I have first shown that the so-called Coulomb string tension is not 
related to the temporal but to the spatial string tension. This relation explains the finite-temperature behavior of the Coulomb string tension, namely the fact that it does not disappear but even increases above the deconfinement transition. I have then studied the quark sector of QCD in Coulomb gauge using a Slater determinant ansatz for the quark wave functional, which includes in particular the quark-gluon coupling by two different Dirac structures. Our calculations show that there is no spontaneous breaking of chiral symmetry when the (linearly rising) infrared part of the Coulomb potential is excluded. Furthermore, choosing the Coulomb string tension from the lattice data we can reproduce the phenomenological value of the quark condensate when the coupling of the quarks to the transverse gluon is included.

I have then extended the Hamiltonian approach to QCD in Coulomb gauge to finite temperatures by compactifying a spatial dimension. Within this approach I have calculated the chiral and dual quark condensate as function of the temperature. Using our zero temperature solution for the quark and gluon sector as input these calculations predict pseudo-critical temperatures of $T_{\chi}^{\mathrm{pc}} \simeq 170 \, \mathrm{MeV}$ for the chiral and $T_{\mathrm{C}}^{\mathrm{pc}} \simeq 198 \, \mathrm{MeV}$ for the deconfinement transition. Within this approach also the Polyakov loop was calculated \cite{Reinhardt:2013iia} and the correct order of the phase transition was found for SU(2) and SU(3). In all these finite-temperature calculations the zero-temperature variational solutions were used as input, which is likely the reason that the critical temperatures currently obtained are too high 
as compared to lattice data. The solution of the variational principle at finite temperature will be the next step in our investigation of the QCD phase diagram.

\section*{Acknowledgement}

This work was supported in part by DFG-RE856/9-2 and by DFG-RE856/10-1.

\bibliography{QCDT0}

\end{document}